\documentclass[acmsmall,nonacm]{acmart}

\usepackage[magicref,omit,nocompress,notitlecaps]{mttex}
\setcopyright{none}
\usepackage{multicol}
\usepackage{xspace}
\usepackage{enumitem}
\usepackage{tikz-cd}
\hypersetup{
  linkcolor = blue
}
\renewcommand{\todo}[1]{}
\input{defs}

\title{Dependent-Type-Preserving Memory Allocation}

\author{Paulette Koronkevich}
\orcid{0000-0003-0325-3305}
\affiliation{%
  \institution{University of British Columbia}
  \city{Vancouver}
  \country{Canada}
}
\email{pletrec@cs.ubc.ca}

\author{William J. Bowman}
\orcid{0000-0002-6402-4840}
\affiliation{%
  \institution{University of British Columbia}
  \city{Vancouver}
  \country{Canada}
}
\email{wjb@williamjbowman.com}
\authorsaddresses{}

\begin{document}

{
  \maketitle
  \makeatletter \gdef\@ACM@checkaffil{} \makeatother
  \section{Introduction}

Dependently typed programming languages such as Coq, Agda, Idris, and F*,
allow programmers to write detailed specifications of their programs
and prove their programs meet these specifications.
However, these specifications can be violated during compilation since they
are erased after type checking.
External programs linked with the compiled program can violate the
specifications of the original program and change the behavior of
the compiled program---even when compiled with a verified compiler.
For example, since Coq does not allow explicitly allocating memory, a
programmer might link their Coq program with a C program that
can allocate memory.
Even if the Coq program is compiled with a verified compiler like CertiCoq
\cite{anand2017}, the external C program can still violate the memory-safe
specification of the Coq program by providing an uninitialized pointer to
memory.
This error could be ruled out by type checking in a language expressive
enough to indicate whether memory is initialized versus uninitialized.
Linking with a program with an uninitialized pointer could be considered
ill-typed, and our linking process could prevent linking with ill-typed
programs.
To facilitate type checking during linking, we can use
\emph{type-preserving compilation}, which preserves the types through
the compilation process.

%Towards this goal, two major passes of a dependent-type-preserving compiler have been developed:
%CPS \cite{bowman2018:cps-sigma} and closure conversion \cite{bowman2018:cccc}.
%Two major passes remain to be developed to obtain a full model of a dependent-type-preserving compiler,
%following the model of the type-preserving System F to Typed Assembly Language (TAL)
%compiler developed by \citet{morrisett1998:ftotal}.
%The two remaining passes are \emph{memory allocation}
%(allocating pairs and closures on the heap) and \emph{code generation}.

In this ongoing work, we develop a typed intermediate language (IL) \tlang
that supports dependent memory allocation, as well as a
dependent-type-preserving compiler pass for memory allocation.
The dependent-type-preserving compiler pass allocates heap space for
dependent pairs and closures from the \slang language developed by
\citet{bowman2018:cccc}.
This work is a significant step towards developing a full
dependent-type-preserving compiler.
In particular, combining this pass with previous work on two major passes,
CPS \cite{bowman2018:cps-sigma} and closure conversion
\cite{bowman2018:cccc}, a dependent-type-preserving compiler to a
low-level language like C is possible. 

  \newcommand{\FigTrans}[1][t]{
  \begin{figure}[#1]
    \judgshape{\all{\se} = \te}
    \begin{displaymath}
      \begin{array}{rcl}
        \all{\ssigmaty{\sx}{\sA}{\sB}} & \defeq & \tsigmaty{\tx}{\init{\all{\sA}}{1}}{\init{\all{\sB}}{1}}
        \\

        \all{\sdpaire{\seone}{\setwo}{\ssigmaty{\sx}{\sA}{\sB}}} & \defeq &
        \begin{stackTL}\talete{\ty}{\tmalloce{\tx}{\all{\sA}}{\all{\sB}}}{\tsigmaty{\tx}{\init{\all{\sA}}{0}}{\init{\all{\sB}}{0}}}{\\\talete{\tyone}{\tassonee{\ty}{\all{\seone}}}{\tsigmaty{\tx}{\init{\all{\sA}}{1}}{\init{\all{\sB}}{0}}}{\\\talete{\tytwo}{\tasstwoe{\tyone}{\all{\setwo}}}{\tsigmaty{\tx}{\init{\all{\sA}}{1}}{\init{\all{\sB}}{1}}}{\\\tytwo}}\end{stackTL}}
          \\
      \all{\scloe{\seone}{\setwo}{\spity{\sx}{\sA}{\sB}}} & \defeq &
      \begin{stackTL}\tlete{\ty}{\tmalloce{\tx}{\all{\scodety{\sxpr:\sAone,\sx:\sA}{\sB}}}{\all{\sAone}}}{\\ \talete{\tyone}{\tassonee{\ty}{\all{\seone}}}{\tsigmaty{\tx}{\init{\all{\scodety{\sxpr:\sAone,\sx:\sA}{\sB}}}{1}}{\init{\sAone}{0}}}{\\ \talete{\tytwo}{\tasstwoe{\tyone}{\all{\setwo}}}{\tsigmaty{\tx}{\init{\all{\scodety{\sxpr:\sAone,\sx:\sA}{\sB}}}{1}}{\init{\sAone}{1}}}{\\\tctage{\tytwo}}}\end{stackTL}}
      \end{array}
    \end{displaymath}
    \vspace{-3ex}
    \caption{Allocation Translation (excerpt)}
    \label{fig:alloc:trans}
    \vspace{-3ex}
  \end{figure}
}

\newcommand{\FigCCCCATyping}[1][t]{
  \begin{figure}[#1]
    \judgshape{\ttyjudg{\thenv}{\tlenv}{\te}{\tA}}
    \begin{mathpar}
      \inferrule[\rulename{Clo}]
      {\ttyjudg{\thenv}{\tlenv}{\te}{\tsigmaty{\ty}{\init{(\tcodety{\txone:\tAone,\tx:\tA}{\tB})}{1}}{\init{\tAone}{1}}}}
      {\ttyjudg{\thenv}{\tlenv}{\tctage{\te}}{\tpity{\tx}{\subst{\tA}{\tsnde{\te}}{\txone}}{\subst{\tB}{\tsnde{\te}}{\txone}}}}

        \inferrule[\rulename{Malloc}]
        {\ttyjudg{\thenv}{\tlenv}{\tA}{\tU} \\
          \ttyjudg{\thenv}{\tlenv,\tx:\tA}{\tB}{\tU}
        }
        {\ttyjudg{\thenv}{\tlenv}{\tmalloce{\tx}{\tA}{\tB}}{\tsigmaty{\tx}{\init{\tA}{0}}{\init{\tB}{0}}}}

        \inferrule[\rulename{Assign1}]
        {\ttyjudg{\thenv}{\tlenv}{\te}{\tsigmaty{\tx}{\init{\tA}{0}}{\init{\tB}{\tphi}}} \\
          \ttyjudg{\thenv}{\tlenv}{\tepr}{\tA}
        }
        {\ttyjudg{\thenv}{\tlenv}{\tassonee{\te}{\tepr}}{\tsigmaty{\tx}{\init{\tA}{1}}{\init{\tB}{\tphi}}}}

        \inferrule[\rulename{Assign2}]
        {\ttyjudg{\thenv}{\tlenv}{\te}{\tsigmaty{\tx}{\init{\tA}{1}}{\init{\tB}{0}}} \\
          \ttyjudg{\thenv}{\tlenv}{\tepr}{\subst{\tB}{\tfste{\te}}{\tx}}
        }
        {\ttyjudg{\thenv}{\tlenv}{\tasstwoe{\te}{\tepr}}{\tsigmaty{\tx}{\init{\tA}{1}}{\init{\tB}{1}}}}

        \inferrule[\rulename{Fst}]
        {\ttyjudg{\thenv}{\tlenv}{\te}{\tsigmaty{\tx}{\init{\tA}{1}}{\init{\tB}{\tphi}}}}
        {\ttyjudg{\thenv}{\tlenv}{\tfste{\te}}{\tA}}

        \inferrule[\rulename{Snd}]
        {\ttyjudg{\thenv}{\tlenv}{\te}{\tsigmaty{\tx}{\init{\tA}{1}}{\init{\tB}{1}}}}
        {\ttyjudg{\thenv}{\tlenv}{\tsnde{\te}}{\subst{\tB}{\tfste{\te}}{\tx}}}
    \end{mathpar}
    \vspace{-2ex}
    \caption{\tlang Typing (excerpt)}
    \label{fig:cc-cca:type}
    \vspace{-5ex}
  \end{figure}
}

\section{Typed memory allocation}

%The typed memory allocation pass presented in this section compiles \slang \cite{bowman2018:cccc},
%the target language from dependent-type-preserving closure conversion, to \tlang, a language
%with explicit memory allocation and initialization operators.
%The dependent pair type also includes initialization tags to indicate whether
%elements have been initialized or not (as is done
%by \citet{morrisett1998:ftotal}). 

%\subsection{The Source Language CC-CC}
The source language \slang of this translation is based on the Calculus of Constructions (CC);
however, first-class functions are replaced by closed code and closures.
\slang includes one impredicative universe \im{\sstarty}, and one predicative
universe \im{\sboxty}.
Expressions have no explicit distinction between terms,
types, or kinds, but we use the meta-variable \im{\se} to evoke a
term expression and \im{\sA} or \im{\sB} to evoke a type expression.
The syntax of expressions include a unit value \im{\sunite} and its
type \im{\sunitty}, let expressions \im{\salete{\sx}{\se}{\sA}{\sepr}},
closed code \im{\snfune{(\sn : \sApr,\sx : \sA)}{\se}} and
dependent code types \im{\scodety{\sn : \sApr,\sx : \sA}{\sB}}, closure
values \im{\scloe{\se}{\se}{\spity{\sx}{\sA}{\sB}}} and dependent closure types
\im{\spity{\sx}{\sA}{\sB}}, application \im{\sappe{\se}{\sepr}}
(which applies closures instead of functions), dependent pairs
\im{\sdpaire{\se}{\sepr}{\ssigmaty{\sx}{\sA}{\sB}}} and their type \im{\ssigmaty{\sx}{\sA}{\sB}}, and finally first and
second projection, \im{\sfste{\se}} and \im{\ssnde{\se}}.
The typing, subtyping, and conversion rules for \slang are the same as given
by \citet{bowman2018:cccc}.

%\subsection{The Target Language}
The target language \tlang includes all the features of \slang,
extending the syntax with an explicit dependent memory allocation
\im{\tmalloce{\tx}{\tA}{\tB}}, and memory initialization operators
\im{\tassonee{\te}{\tepr}} and \im{\tasstwoe{\te}{\tepr}}.
\tlang only allocates tuples of two machine words; this suffices for our
translation, but it should not be difficult to extend the language
to arbitrary length tuples.
The type of dependent pairs
\im{\tsigmaty{\tx}{\init{\tA}{\tphi}}{\init{\tB}{\tphi}}}
includes an initialization flag \im{\tphi} to indicate whether
the elements have been initialized (\im{1}) or not (\im{0}),
as is done by \citet{morrisett1998:ftotal}.
This prevents the first and second projection from accessing
unallocated memory.
Finally, since closures are allocated as pairs, we use a tag
\im{\tctage{\te}} to indicate the pair should be treated as a closure.

The typing rules for \tlang remain mostly the same as \slang; however,
we now include the heap while type checking to access tuples
that have been allocated.
The heap \im{\thenv} consists of locations with their types
\im{\tfl:\tA} as well as locations mapped to heap values
\im{\locmap{\tfl}{\tpaire{\teone}{\tetwo}}}.
Small step reduction \im{\step} and conversion \im{\stepstar} are
defined over configurations \im{\conf{\thenv}{\te}} to access
the allocated tuples.
For example, first and second projection of a pair are defined as follows:
\begin{displaymath}
  \vspace{-1ex}
      \begin{array}{rcll}
       \conf{\thenv}{\tfste{\tfl}} & \step_{\pi_{1}} & \conf{\thenv}{\teone} & \where{\locmap{\tfl}{\tpaire{\teone}{\tetwo}} \in \thenv} \\
       \conf{\thenv}{\tsnde{\tfl}} & \step_{\pi_{2}} & \conf{\thenv}{\tetwo}
       & \where{\locmap{\tfl}{\tpaire{\teone}{\tetwo}} \in \thenv}
      \end{array}
\end{displaymath}

\FigCCCCATyping
The typing rules for memory allocation, initialization, closure
tagging, and first and second projection
are given in  \fullref[]{fig:cc-cca:type}.
Memory allocation, as expected, types as a dependent pair with the
initialization flags set to \im{0}.
Memory initialization changes the flag on the dependent pair type
depending on which element is initialized.
Finally, \rulename{Clo} types a dependent pair as a \tfonttext{\im{\Pi}} type,
given the elements of the pair are closed code and its environment. 

%\subsection{The Translation}
\FigTrans
The essence of the translation is given in \fullref[]{fig:alloc:trans}.
All other cases simply recursively translate subexpressions.
For pairs and closures, the translated program first initializes
memory according to the type of the dependent pair (or the closure).
Then, the program initializes each element of the pair with the result of
translating the subexpressions \im{\seone} and \im{\setwo}.
Finally, it returns the result of this allocation, tagging it as a closure
in the closure case. 

  \section{Type preservation}

Type preservation guarantees that a well-typed source program is compiled to
a well-typed target program, and is stated as follows:
\begin{theorem}
  If \im{\styjudg{\slenv}{\se}{\sA}} then
  \im{\ttyjudg{\cdot}{\all{\slenv}}{\all{\se}}{\all{\sA}}}.
\end{theorem}
\noindent Since typing in dependently typed languages relies on subtyping and
equivalence, and equivalence relies on conversion, we must also
prove that subtyping, equivalence, and conversion are preserved by the
allocation pass:
\begin{lemma}
  If \im{\ssubtyjudg{\slenv}{\sA}{\sB}} then
  \im{\tsubtyjudg{\cdot}{\all{\slenv}}{\all{\sA}}{\all{\sB}}}.
\end{lemma}
\begin{lemma}
  If \im{\sequivjudg{\slenv}{\se}{\sepr}} then
  \im{\tequivjudg{\cdot}{\all{\slenv}}{\all{\se}}{\all{\sepr}}}.
\end{lemma}
\begin{lemma}
  If \im{\sstepjudg{\slenv}{\se}{\sepr}} then
  \im{\tstepjudg{\cdot}{\all{\slenv}}{\conf{\cdot}{\all{\se}}}{\conf{\thenv}{\all{\sepr}}}}.
\end{lemma}

\noindent Finally, since many dependent typing rules substitute terms into types
(such as \rulename{Snd}), we must also show that substitution is
preserved: 
\begin{lemma}
  \im{\all{\subst{\se}{\sepr}{\sx}} \equiv \subst{\all{\se}}{\all{\sepr}}{\tx}}.
\end{lemma}

The proofs of all these lemmas and the main type preservation theorem follow
by straightforward induction over the source derivation.
Intuitively, the translation is only adding series of \tfonttext{let}
statements which explicitly allocate a pair (or closure).
The final compiled program should be a pair (or closure) of the same type.

  \section{Future Work}
Proving type preservation to an arbitrary target language is not enough,
as the target language itself should be type-safe and consistent
(that is, we cannot prove false).
We prove \tlang to be type-safe and consistent by providing a model in
extensional CIC (eCIC).
We then prove the model preserves typing and the definition of the
empty type (\im{\bot}), following a technique well explained
by \citet{boulier2017}.
If \tlang was inconsistent, then we could produce a proof of false
\tfonttext{\im{\bot}} and translate it to \im{\bot} in eCIC;
however, since no such proof exists in eCIC, \tlang must be consistent.

We translate each dependent pair to a dependent pair in eCIC, where each
element has type \im{\mmaybe{A}}, with a proof about which element of
the pair is filled based on the initialization flags:
\begin{displaymath}
  \vspace{-1ex}
      \begin{array}{rcl}
        \model{\tsigmaty{\tx}{\init{\tA}{0}}{\init{\tA}{0}}}  & \defeq &
        \msigmaty{\mp}{(\msigmaty{\mx}{\mmaybe{\model{\tA}}}{\mmaybe{\model{\tB}}})}{\mp = \mpaire{\mnone}{\mnone}} \\
        \model{\tsigmaty{\tx}{\init{\tA}{1}}{\init{\tA}{0}}}  & \defeq &
        \msigmaty{\mp}{(\msigmaty{\mx}{\mmaybe{\model{\tA}}}{\mmaybe{\model{\tB}}})}{\mp = \mpaire{\mjust{\me}}{\mnone}} \\
        \model{\tsigmaty{\tx}{\init{\tA}{1}}{\init{\tA}{1}}}  & \defeq &
        \msigmaty{\mp}{(\msigmaty{\mx}{\mmaybe{\model{\tA}}}{\mmaybe{\model{\tB}}})}{\mp = \mpaire{\mjust{\meone}}{\mjust{\metwo}}}
      \end{array}
\end{displaymath}

\noindent For readibility we exclude the existential quantification
for each element inside Just expressions.
We then add auxiliary definitions to eCIC, maybe-snd and maybe-fst,
which are guaranteed to produce the first and second element of a pair,
since they expect a pair alongside a proof that the elements are filled:
\begin{displaymath}
    \vspace{-1ex}
      \begin{array}{lcl}
  \text{maybe-fst} : \msigmaty{\mp}{(\msigmaty{\mx}{\mmaybe{\mA}}{\mmaybe{\mB}})}{\mp = \mpaire{\mjust{\me}}{\_}} \rightarrow \mA
  \\
  \text{maybe-snd} : \msigmaty{\mp}{(\msigmaty{\mx}{\mmaybe{\mA}}{\mmaybe{\mB}})}{\mp = \mpaire{\mjust{\meone}}{\mjust{\metwo}}}\rightarrow \subst{\mB}{\meone}{\mx}
  \end{array}
\end{displaymath}

\noindent Any instances of \tfonttext{fst} and \tfonttext{snd} are then
translated to maybe-fst and maybe-snd.
Based on \rulename{Fst} and \rulename{Snd}, the translation of \im{\te}
should have the type expected by these definitions,
so the model should be type-preserving.
Proving this translation to eCIC type preserving is still ongoing.

An exciting extension of \tlang would be
an explicit \tfonttext{free} operator to deallocate memory.
%This operator allows fine-grained control over memory, as well
%as the ability to safely compile-then-link Coq components with
%components written or compiled to this IL.
This IL with \tfonttext{free} would be an interesting language to
implement a garbage collector for the dependent-type-preserving
compiler model so far, giving us even further confidence in the
correct evaluation of dependently-typed programs.
This IL would likely need richer specifications of memory operations,
like those provided by Hoare Type Theory \cite{nanevski2006}.

%\subsection{Adding a free operator}

  \bibliographystyle{ACM-Reference-Format}
  \bibliography{paulette}

}

\end{document}